\documentclass[]{spie}  

\usepackage{amsmath,amsfonts,amssymb}
\usepackage{bm}
\usepackage{graphicx}
\usepackage[colorlinks=true, allcolors=blue]{hyperref}
\usepackage{xcolor}
\usepackage{subcaption}
\usepackage[perpage]{footmisc} 
\usepackage{tabularx}
\usepackage{cite}
\usepackage{wrapfig}
\title{Optimizing Wavefront-Deformation Sensor Placement for Active
Radio-Telescope Surfaces}

\author[a]{Stefan Thoms}
\author[a]{Martin Timpe}
\author[a]{Matthias Reichert}
\affil[a]{OHB Digital Connect GmbH, Weberstrasse 21, 55130 Mainz, Germany}

\authorinfo{\hspace*{-0.2cm}Send correspondence to stefan.thoms@ohb.de}

\begin{document} 
\maketitle

\begin{abstract}
Next-generation high-frequency radio telescopes require primary-surface
accuracies that passive structures alone cannot reliably achieve. The Atacama
Large Aperture Submillimeter Telescope (AtLAST\footnote[2]{\href{https://www.atlast.uio.no/}{https://www.atlast.uio.no/}}),
a 50\,m single-dish concept operating up to $\approx$\,1\,THz, imposes a
$\approx$\,20\,\textmu m\,rms surface-accuracy requirement across its full
aperture. This is practically unattainable for a purely passive reflector
subject to gravitational, thermal, and wind-induced deformation. Closed-loop active collimation and surface control are therefore imperative, which in turn requires the deformation field to be known across the full aperture in real time. Measuring it directly at the necessary resolution across the complete surface is, however, hardly feasible; instead, the current AtLAST concept development foresees reconstructing the surface from a limited set of discrete sensor positions.

An algorithmic framework is presented that optimizes the number and placement of
these sensors to maximize the reconstructability of the deformation field.
Finite-element analysis (FEA) load cases (gravity, thermal, wind) define the
deformation space, from which a data-driven Proper Orthogonal Decomposition
(POD) basis is derived; sensor positions are then chosen by a greedy
optimization algorithm and then assessed via leave-one-out cross-validation.
Applied to FEA deformations of AtLAST's Back-Up Structure (BUS), and assuming a
sensor noise of 5\,µm\,rms, the method reconstructs all load
cases with 50 sensors to below 2.7\,\textmu m\,rms (worst case) residual (BUS-)
surface error. 

\end{abstract}


\section{INTRODUCTION}
\label{sec:intro}
The Atacama Large Aperture Submillimeter Telescope (AtLAST) is a 50\,m
single-dish facility concept for \mbox{(sub-)millimeter} astronomy.\cite{mroczkowski2025conceptual, cicone2026atacama, Klaassen2020}
Operating at 30--950\,GHz with a field of view of up to 2$^\circ$, AtLAST will
map the submillimeter sky $10^3$--$10^5$ times faster than ALMA while
maintaining comparable point-source sensitivity.\cite{mroczkowski2025conceptual, Cicone2026, Reichert2024}
Observations at the highest frequencies require a half-wavefront error (HWFE)
of $\approx$\,20\,\textmu m\,rms, according to a Ruze efficiency
calculation.\cite{mroczkowski2025conceptual, Reichert2024}

\begin{wrapfigure}{l}{0.32\columnwidth}
  \centering
  \vspace{-\baselineskip}
  \includegraphics[width=\linewidth]{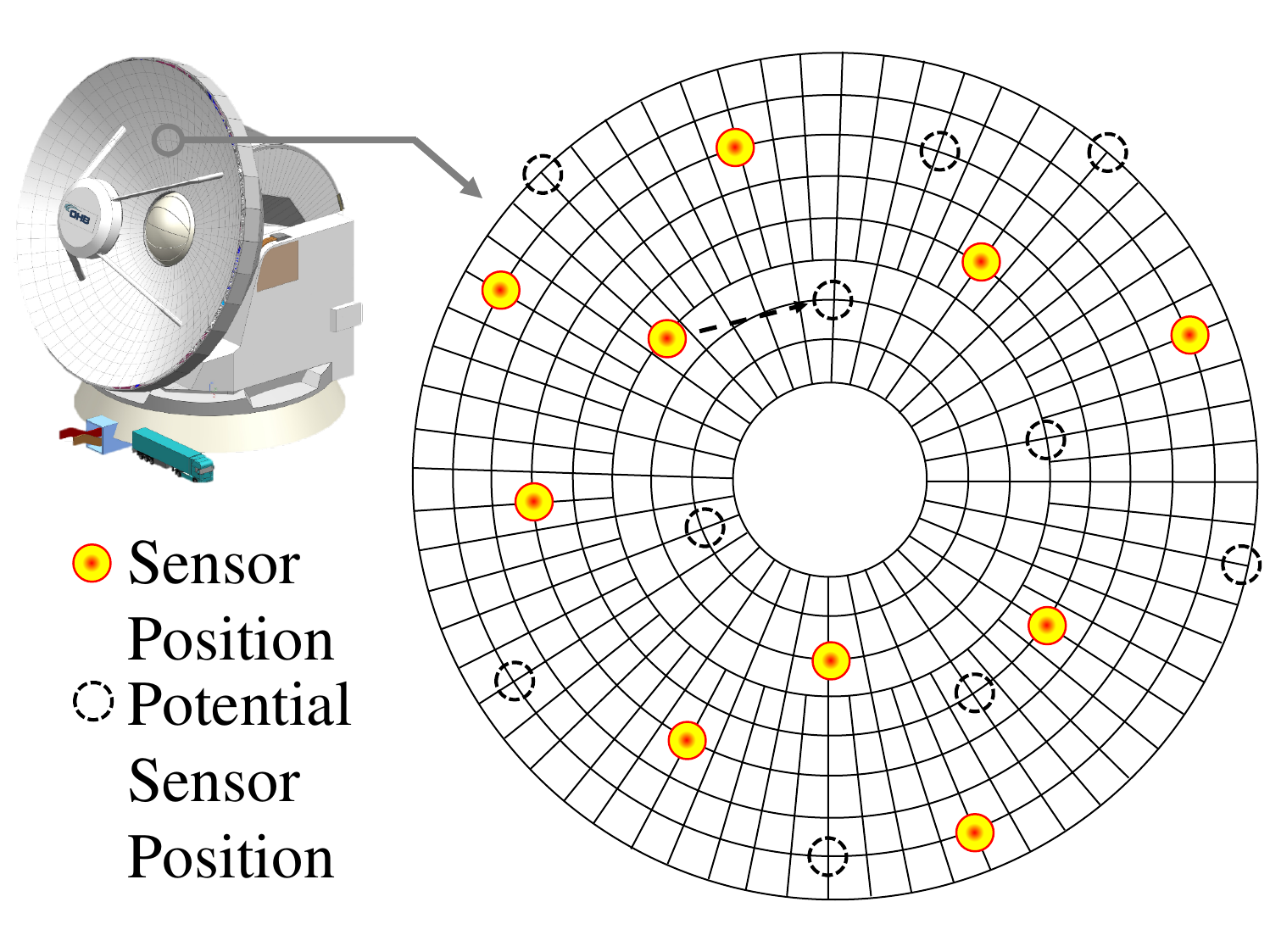}
  \setlength{\abovecaptionskip}{-6pt}
  \caption{Sensors on AtLAST's M1.}  
  \label{fig:intro}
  \vspace{-\baselineskip}
\end{wrapfigure}

Finite-element analysis (FEA) of AtLAST's structure shows that a passive
reflector of this size falls well short of the $\approx$\,20\,\textmu m\,rms
surface-accuracy requirement.\cite{Reichert2024} Achieving it
demands an active primary surface and collimation that compensate for these
deformations in closed loop, which in turn requires the deformation field to
be known across the full dish in real time.

Regardless of the sensing technology employed, measuring the deformation field in real time across the complete surface has no practical solution to date. The surface must therefore be reconstructed from a limited number of point measurements. The deformations of interest, however, are dominated by low
spatial orders set by the structural topology of the Back-Up-Structure (BUS), which makes such reconstruction viable, provided the sensors are positioned where they are most informative.
We present a framework that addresses this problem. FEA load cases define the deformation space of the primary reflector; a Proper Orthogonal Decomposition (POD) basis is extracted from these snapshots; sensor positions are selected by greedy optimal placement and the configuration is validated by leave-one-out cross-validation (LOOCV).
The method is demonstrated on the AtLAST BUS but applies to any structure for which representative load cases can be computed. 
\\

\section{METHODOLOGY}
\label{sec:method}
The framework, implemented as a Python toolbox, consists of five sequential stages: data input, sensor positioning, sensor readout, surface reconstruction, and performance evaluation. Each stage is described below in the order of the processing pipeline.

\subsection{Data input}
\label{sec:data_input}
\begin{figure}[t]
  \centering
  \includegraphics[width=1\columnwidth]{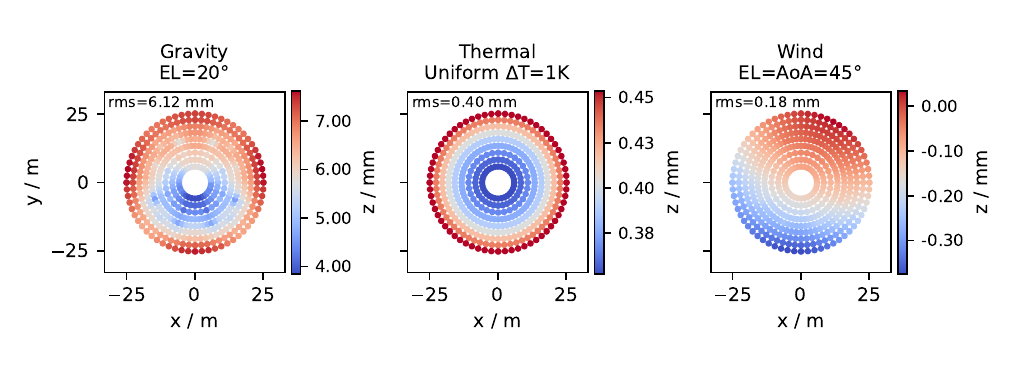}
  \caption{Surface deformation from FEA for selected structural load cases:
  gravitational (EL\,=\,20\textdegree, left); thermal, $\Delta T$\,=\,1\,K
  applied uniformly to the structure (middle); and wind at 10\,m/s,
  EL\,=\,AoA\,=\,45\textdegree{} (right).}
  \label{fig:fea_load_cases}
\end{figure}

The input to the framework is a set of $M$\,=\,100 deformation load cases for
AtLAST computed by finite-element analysis (FEA). These comprise six
gravitational cases spanning elevation angles EL\,=\,20\,--\,90\textdegree;
16 idealized thermal cases, each applying either a uniform $\Delta
T$\,=\,$+1$\,K to a complete (sub-)system or a $\Delta T$\,$\approx$\,1\,K
gradient across (sub-)systems; and 78 wind cases at 10\,m/s, covering six
elevation angles and 13 angles of attack (AoA). Figure~\ref{fig:fea_load_cases}
shows one representative case from each category.

Each load case provides the displacements relative to the nominal geometry at
$P$\,=\,480 grid points, forming a column of the snapshot matrix
\begin{equation}
    \bm{S}_\mathrm{FEA} \;\in\; \mathbb{R}^{P \times M}.
    \label{eq:snapshot_matrix}
\end{equation}
The grid points are the nodes of the FEA mesh representing the back-up structure (BUS) on which the reflector panels are mounted. The evaluation therefore characterizes the deformation and reconstruction of the BUS geometry alone, not the complete surface error, to which further contributors (e.g., panel surface accuracy and M2) would need to be added.

As an alternative to FEA data, real measurements from photogrammetry or holography could serve as input, given sufficient accuracy and a representative diversity of load cases.

\subsection{Sensor positioning}
\label{sec:sensor_positioning}
Sensor placement is basis-aware. The primary basis used is a data-driven basis extracted by Proper Orthogonal Decomposition (POD), derived directly from the FEA load cases of (Sec.~\ref{sec:data_input}); for comparison, an analytic basis of annular Zernike polynomials is also evaluated. In both cases the basis is truncated to $k$ modes, and $N$ sensor locations are then determined by greedy selection on the chosen basis. FEA grid nodes coincide with panel-gap locations on the BUS  and thus form physically reasonable candidate positions; both strategies select from this node set.

\paragraph{Annular Zernike basis.}
The annular Zernike polynomials are evaluated on the FEA grid.\cite{Mahajan1981}
Each polynomial is characterized by its radial order $n$ and azimuthal
frequency $m$; here, the modes are enumerated by the Noll single-index
convention $j$, which orders them by increasing radial order.\cite{Noll1976}
A truncation to $k$ modes thus retains the first $k$ polynomials of this
sequence; Table~\ref{tab:zernike_modes} lists the leading modes. Piston
($j$\,=\,1) is retained to represent a constant offset across the aperture.
Standard Zernike polynomials lose orthogonality on annular domains;
orthonormality is therefore restored numerically via QR decomposition on the
discrete point cloud.\cite{Upton2004} Being analytic, this basis is fixed and
independent of the load-case library.
\begin{table}[h]
  \centering
  \caption{Leading Zernike modes in Noll ordering.}
  \label{tab:zernike_modes}
  \setlength{\tabcolsep}{16pt}   
  \begin{tabular}{ccl}
    \hline
    $j$ & $(n,\,m)$ & Name \\
    \hline
    1  & $(0,\,0)$  & Piston \\
    2  & $(1,\,1)$  & Tilt $x$ \\
    3  & $(1,\,-1)$ & Tilt $y$ \\
    4  & $(2,\,0)$  & Defocus \\
    5  & $(2,\,-2)$ & Oblique astigmatism \\
    6  & $(2,\,2)$  & Vertical astigmatism \\
    7  & $(3,\,-1)$ & Vertical coma \\
    8  & $(3,\,1)$  & Horizontal coma \\
    9  & $(3,\,-3)$ & Oblique trefoil \\
    10 & $(3,\,3)$  & Vertical trefoil \\
    11 & $(4,\,0)$  & Primary spherical \\
    ... & ...  & ...\\
    \hline
  \end{tabular}
\end{table}

\paragraph{POD basis.}
The snapshot matrix is first mean-centered by subtracting the per-node mean across all load cases. Its singular value decomposition (SVD),
\begin{equation}
\bm{S}_\text{FEA} = \bm{U}\, \bm{\Sigma}\, \bm{V}^{\!\top},
\label{eq:svd}
\end{equation}
yields left singular vectors $\bm{U}$ (spatial modes) and singular values $\sigma_i$ on the diagonal of $\bm{\Sigma}$, ranking the modes by significance. The POD basis $\bm{U}_k \in \mathbb{R}^{P \times k}$ retains the leading $k$ columns of $\bm{U}$, i.e.\ the modes with the largest singular values. These modes span the subspace best approximating the deformation space sampled by the load-case library. Unlike the fixed Zernike basis, the POD modes adapt to the mechanical properties of the specific structure.

\begin{figure}[h] 
    \centering 
    \includegraphics[width=0.5\columnwidth]{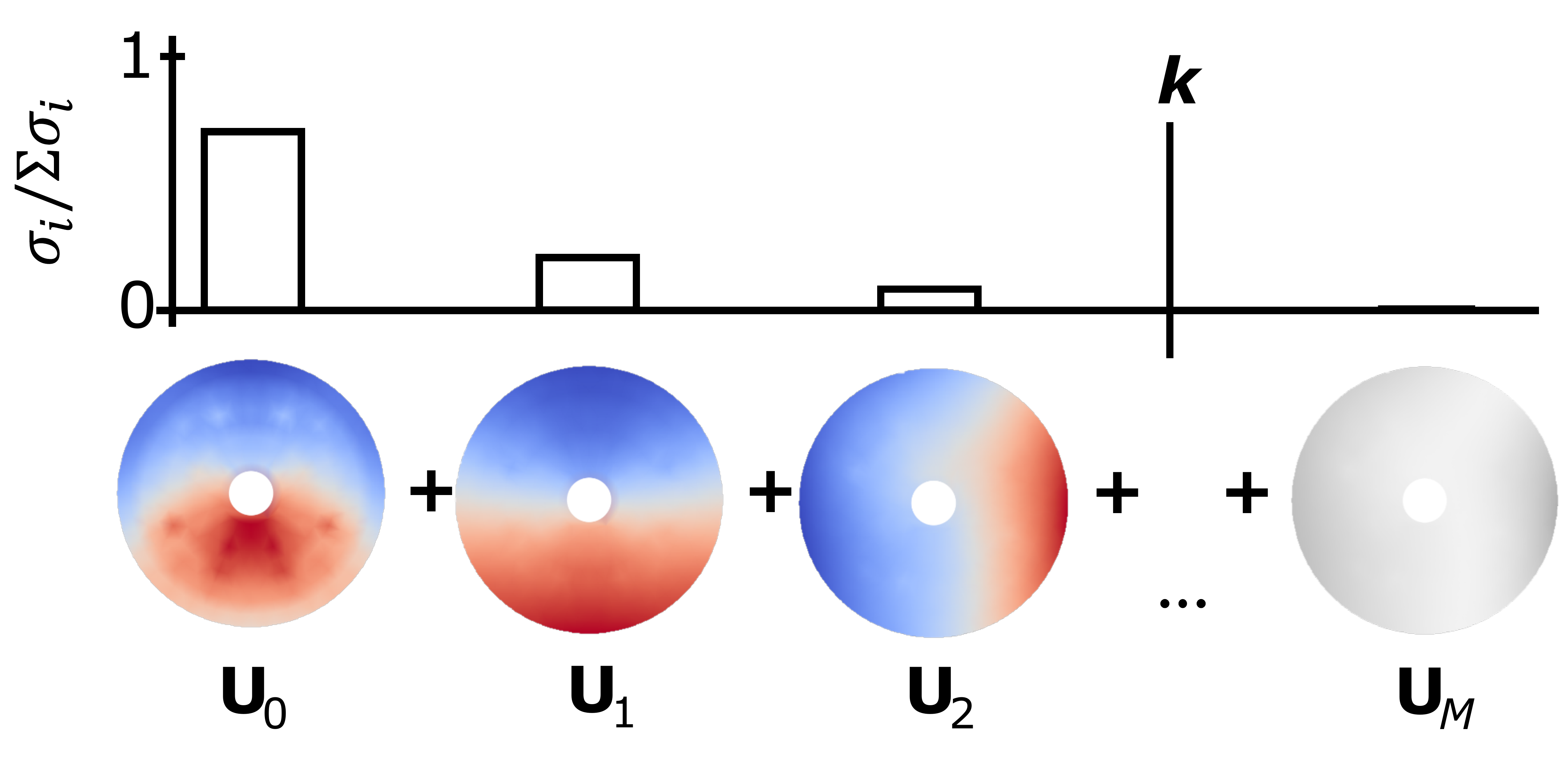} \setlength{\abovecaptionskip}{6pt} 
    \caption{Schematic of the POD basis. Top: illustration of the relative singular-value significance $\sigma_i / \sum_i \sigma_i$, decreasing with mode index. Bottom: the three leading POD modes. The cut at $k$ indicates truncation to the first $k$ modes.} 
    \label{fig:svd_u} 
\end{figure}

\paragraph{Greedy sensor selection.}
The chosen basis is denoted $\bm{\Psi} \in \mathbb{R}^{P \times k}$, corresponding to $\bm{U}_k$ in the POD case and to the annular Zernike matrix in the analytic case. Selecting $N$ sensors amounts to selecting $N$ rows of $\bm{\Psi}$, defining the sub-matrix $\bm{\Psi}_N \in \mathbb{R}^{N \times k}$. Each candidate set defines an information matrix $\bm{\Psi}_N^\top \bm{\Psi}_N$ that quantifies how well the $k$ basis coefficients can be recovered. 

Different standard criteria can be applied on this matrix for the optimization: D-optimality maximizes its determinant (mode distinguishability), I-optimality minimizes the aperture-averaged prediction variance (uniform accuracy), and G-optimality minimizes the worst-case variance (bounded local error). As analysis showed, that results do not differ too much in terms of the optimization criterium, I-optimality is used as a default to achieve most uniform accuracy. Since identifying the globally optimal subset is combinatorial, sensors are instead selected greedily~\cite{Manohar2018}: at each step, the candidate that most improves the criterion is appended, until $N$ sensors are selected. The resulting greedy set is then refined via pairwise swaps~\cite{Fedorov1972}, in which each selected sensor is exchanged with the best non-selected candidate whenever doing so improves the criterion (100 iterations).
\\

\subsection{Sensor readout}
\label{sec:sensor_readout}
Sensor readings are simulated by sampling the FEA deformation field at the selected sensor locations and adding measurement noise:
\begin{equation}
\bm{d}_j = \bm{C}\, \bm{s}_j + \bm{n}, \qquad \bm{n} \sim \mathcal{N}(\bm{0},\, \sigma^2 \bm{I}),
\label{eq:readout}
\end{equation}
where $\bm{s}_j \in \mathbb{R}^{P}$ is the deformation of load case $j$ (one column of $\bm{S}_\text{FEA}$), $\bm{C} \in \{0,1\}^{N \times P}$ is the Boolean selection matrix picking the $N$ sensor rows, and $\bm{n} \in \mathbb{R}^{N}$ is a vector of independent zero-mean Gaussian noise samples of standard deviation $\sigma$ (one per sensor); $\sigma = 5\,$\textmu m (1$\sigma$) is used throughout.\\
%
%
\subsection{Surface reconstruction}
\label{sec:reconstruction}
Given the sensor readout $\bm{d}_j \in \mathbb{R}^N$ and the basis matrix $\bm{\Psi}$, the deformation field is reconstructed by fitting basis coefficients to the measurements:
\begin{equation}
    \bm{d}_j = \bm{\Psi}_N\,\bm{a}_j + \bm{e}_j,
    \label{eq:sensor_model}
\end{equation}
where $\bm{\Psi}_N \in \mathbb{R}^{N \times k}$ is the sensor-location submatrix of $\bm{\Psi}$ and $\bm{e}_j$ collects the residual not captured by the truncated basis. The coefficients follow from ordinary least squares,
\begin{equation}
    \hat{\bm{a}}_j = \left(\bm{\Psi}_N^\top\,\bm{\Psi}_N\right)^{-1}\bm{\Psi}_N^\top\,\bm{d}_j,
    \label{eq:lstsq}
\end{equation}
evaluated via the SVD-based pseudoinverse. The reconstructed full-field deformation is then
\begin{equation}
    \hat{\bm{s}}_j = \bm{\Psi}\,\hat{\bm{a}}_j \in \mathbb{R}^P.
    \label{eq:reconstruct}
\end{equation}
For the POD basis, the per-node mean deformation is subtracted from $\bm{d}_j$ before fitting and added back to $\hat{\bm{s}}_j$ after expansion, consistent with the mean-centering used during basis extraction.
\\
\\

\subsection{Performance evaluation}
\label{sec:evaluation}
The reconstruction error for load case $j$ is given by the residuals
\begin{equation}
    \bm{r}_j = \bm{s}_j - \hat{\bm{s}}_j,
    \label{eq:residual}
\end{equation}
from which a root-mean-square (rms) surface error is obtained. The primary performance metric is the per-load-case rms error; the framework's target is that every load case is reconstructed below a prescribed rms threshold. For AtLAST, the overall surface accuracy target is 20\,\textmu m. Since the BUS is not representing the overall surface error, the reconstruction accuracy targeted here is set to an ambitious rms $\le$\,5\,\textmu m.

For the POD basis, validation follows a leave-one-out cross-validation (LOOCV) scheme to obtain a more realistic estimate of reconstruction performance. In each fold, one load case $j$ is held out and the POD basis is recomputed from the remaining $M-1$ snapshots, while the sensor positions determined once on the full set are retained to reduce computational effort. The held-out load case is then reconstructed from its simulated sensor readings using the fold-specific basis. This tests reconstruction on genuinely unseen deformations while keeping the sensor geometry fixed. The annular Zernike basis, by contrast, is analytic and independent of the load-case library, so no cross-validation is required; all $M$ load cases are reconstructed directly.
\\
\section{RESULTS}
\label{sec:results}

Both bases are evaluated over sensor counts $N \in \{30,\allowbreak 50,\allowbreak 100,\allowbreak 200,\allowbreak 250\}$ and truncations $k \in \{10,\allowbreak 20,\allowbreak 30\}$ modes for the POD basis and $k \in \{10,\allowbreak 20, \allowbreak 50\}$ modes for the annular Zernike basis. Only configurations with $N > k$ are considered (i.e., number of sensors $N$ is larger than the number of reconstruction modes $k$), so that the least-squares problem in Eq.~\eqref{eq:lstsq} remains
overdetermined. All configurations use greedy I-optimal sensor selection with
subsequent swap refinement (Sec.~\ref{sec:sensor_positioning}) and Gaussian
sensor noise of 5\,\textmu m (1$\sigma$), as defined in
Eq.~\eqref{eq:readout}. POD results are obtained under LOOCV, while Zernike
reconstructions are evaluated directly on all $M$ load cases
(Sec.~\ref{sec:evaluation}).

\subsection{Reconstruction performance}
\label{sec:results_all}

\begin{figure}[t]
  \centering
  \begin{subfigure}[b]{0.48\columnwidth}
    \centering
    \includegraphics[width=\linewidth]{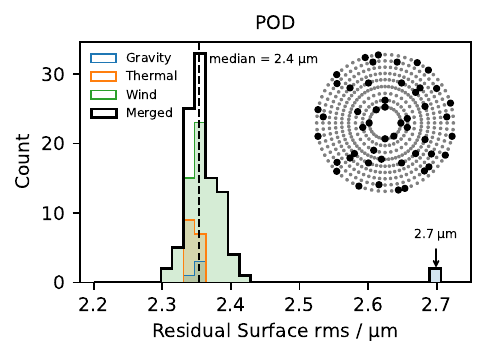}
    \caption{}
    \label{fig:N50_k20_histos_pod}
  \end{subfigure}%
  \hfill
  \begin{subfigure}[b]{0.48\columnwidth}
    \centering
    \includegraphics[width=\linewidth]{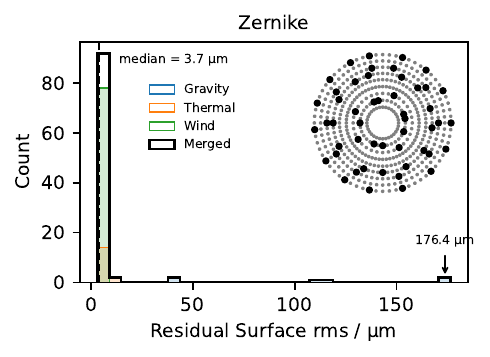}
    \caption{}
    \label{fig:N50_k20_histos_zernike}
  \end{subfigure}
  \caption{Distribution of the per-load-case reconstruction rms for $N$\,=\,50
  sensors, $k$\,=\,20 modes, all 100 load cases, with the corresponding greedy
  I-optimal sensor positions (inset): (a) POD basis under LOOCV, (b) annular
  Zernike basis.}
  \label{fig:N50_k20_histos}
\end{figure}

Figure~\ref{fig:N50_k20_histos} shows the distribution of the per-load-case
reconstruction rms for the configuration $N$\,=\,50, $k$\,=\,20, together with
the corresponding sensor positions (inset). The POD basis reconstructs all
100 load cases to a median of 2.4\,\textmu m and a worst case of 2.7\,\textmu m,
meeting the $\le$\,5\,\textmu m target. The Zernike basis reaches a comparable
median of 3.7\,\textmu m, but its worst case of 176.4\,\textmu m exceeds the
target by more than an order of magnitude. The upper tail of the distribution
consists exclusively of gravitational load cases, indicating that the annular
Zernike basis cannot adequately represent these large-amplitude deformation
patterns at a truncation of $k$\,=\,20 modes.

\begin{figure}[htb]
  \centering
  \includegraphics[width=1\columnwidth]{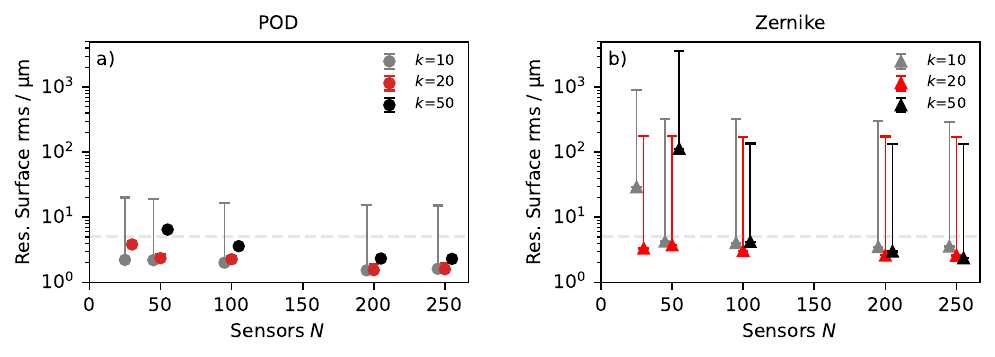}
  \caption{Residual surface rms versus sensor count $N$ for different
  truncations $k$, over all 100 load cases (markers: median; error bars:
  worst case): (a) POD basis under LOOCV, (b) annular Zernike basis. The
  dashed gray line indicates the 5\,\textmu m reconstruction target.}
  \label{fig:performances_inclgrav}
\end{figure}

Figure~\ref{fig:performances_inclgrav} generalizes this observation across the
full sweep. For the POD basis, the residual decreases with $N$ and appears to
converge within a range, where the coefficient estimate becomes
noise-limited. Intuitively, the truncation $k$ further sets an attainable floor,
since modes beyond $k$ cannot be represented regardless of sensor count. The
target is first met at $N$\,=\,30, $k$\,=\,20, including the worst-case load
case. For the Zernike basis, the worst case remains above 5\,\textmu m for all
evaluated $(N, k)$ combinations: the gravitational deformation patterns are
not efficiently representable by low-order annular Zernike polynomials, and
increasing $k$ to 50 does not close the gap.

\subsection{Evaluation excluding gravitational load cases}
\label{sec:results_nograv}

Gravitational deformation differs from the thermal and wind contributions in two respects: its absolute amplitude is far larger (see the representative example in Fig.~\ref{fig:fea_load_cases}), and it is, to a large extend, a systematic and repeatable function of (only) elevation. In principle it can therefore be compensated separately, e.g.\ offline via a look-up table or an elevation-dependent model. The following evaluation repeats the analysis with the six gravitational cases excluded from the performance statistics. The pipeline itself is unchanged: the gravitational cases remain in the snapshot library, and basis extraction and sensor positions are identical to Sec.~\ref{sec:results_all}.

\begin{figure}[b]
  \centering
  \begin{subfigure}[b]{0.48\columnwidth}
    \centering
    \includegraphics[width=\linewidth]{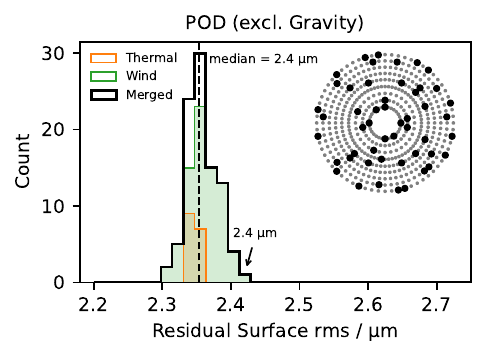}
    \caption{}
    \label{fig:N50_k20_histos_pod_nograv}
  \end{subfigure}%
  \hfill
  \begin{subfigure}[b]{0.48\columnwidth}
    \centering
    \includegraphics[width=\linewidth]{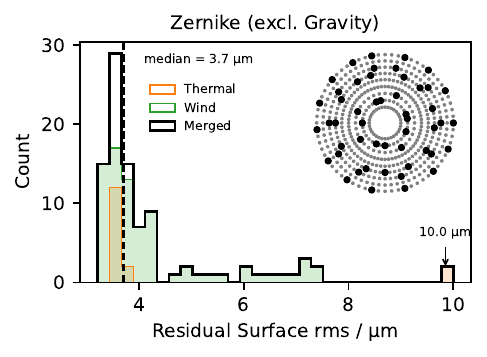}
    \caption{}
    \label{fig:N50_k20_histos_zernike_nograv}
  \end{subfigure}
  \caption{As Fig.~\ref{fig:N50_k20_histos}, but with the six gravitational
  load cases excluded from the evaluation statistics (sensor positions and
  basis unchanged): (a) POD basis under LOOCV, (b) annular Zernike basis.}
  \label{fig:N50_k20_histos_nograv}
\end{figure}

Figure~\ref{fig:N50_k20_histos_nograv} shows the resulting distributions for
$N$\,=\,50, $k$\,=\,20. For the POD basis
(Fig.~\ref{fig:N50_k20_histos_pod_nograv}) the picture is essentially
unchanged relative to Fig.~\ref{fig:N50_k20_histos_pod}: median and worst
case now both round to 2.4\,\textmu m, essentially indistinguishable at this
precision ($\widetilde{\mu}_\text{median}$\,= 2.35\,µm vs. $\mu_\text{max.}$\,= 2.43\,µm), since the excluded gravitational cases were already reconstructed
accurately. For the Zernike basis, in contrast, the worst case drops from
176.4\,\textmu m to 10.0\,\textmu m, consistent with the limited capability
of the basis to represent the gravitational deformation patterns.

\begin{figure}[t]
  \centering
  \includegraphics[width=1\columnwidth]{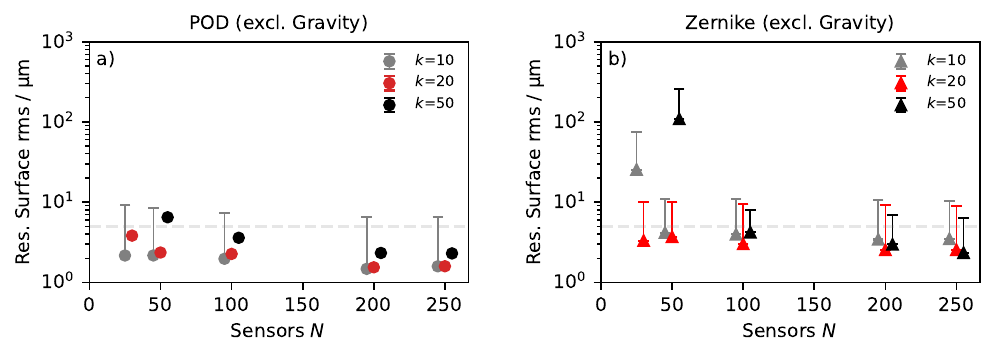}
  \caption{As Fig.~\ref{fig:performances_inclgrav}, but with the six
  gravitational load cases excluded from the evaluation statistics (markers:
  median; error bars: worst case): (a) POD basis under LOOCV, (b) annular
  Zernike basis. The dashed gray line indicates the 5\,\textmu m
  reconstruction target.}
  \label{fig:performances_exclgrav}
\end{figure}

Figure~\ref{fig:performances_exclgrav} shows the corresponding sweep over $N$
and $k$ with the gravitational load cases excluded. The POD curves are nearly
identical to Fig.~\ref{fig:performances_inclgrav}a, again reflecting that
gravity does not drive the POD residuals. For the Zernike basis, both median
and worst case improve substantially: the median falls below 5\,\textmu m
residual surface rms already at moderate $N$, and the worst case also drops
below this threshold once $N > 250$. Only under gravity pre-compensation, the
annular Zernike basis thus could potentially become a viable alternative to the data-driven
POD basis, though a much less efficient one.

\section{SUMMARY AND OUTLOOK}
\label{sec:conclusion}

A framework for optimizing the number and placement of deformation sensors on
active radio-telescope primary surfaces has been presented. From FEA load
cases, a data-driven POD basis is extracted, sensor positions are selected by
greedy I-optimal search with swap refinement, and the configuration is
evaluated by leave-one-out cross-validation. Applied to the AtLAST 50\,m BUS,
the POD basis clearly outperforms an annular Zernike basis, the conventional
analytic choice: with 50 sensors and 20 modes, all 100 load cases are
reconstructed to below 2.7\,\textmu m\,rms, whereas the Zernike basis misses
the target on the gravitational cases by more than an order of magnitude and,
even with gravity excluded, does not meet the 5\,\textmu m target for all load
cases in any evaluated configuration up to $N$\,=\,250 sensors and
$k$\,=\,50 modes.

This capability has practical consequences beyond reconstruction accuracy. If
gravitational deformation is captured by the sensor-based reconstruction,
high-accuracy calibration of the surface (e.g.\ by holography) would, in
principle, be required at a single elevation angle only, rather than across
the full elevation range, substantially reducing the calibration effort. Furthermore, meeting the
reconstruction target with 30--50 sensors instead of several hundred
simplifies the system design, increases robustness, and reduces maintenance
demands, ultimately favoring higher observing efficiency.

Several extensions are planned. The load-case library will be augmented with
additional deformation sources, in particular inertial loads from telescope
acceleration and centrifugal contributions from slewing. Moreover, the present
results are based on single noise realizations per load case; a statistical
evaluation over repeated realizations shall be performed to provide confidence
intervals for the reported residuals.

\acknowledgments 
This project has received funding from the European Union's Horizon Europe research and innovation programme under grant agreement No.\ 101188037 (AtLAST2). Views and opinions expressed are however those of the authors only and do not necessarily reflect those of the European Union or European Research Executive Agency. Neither the European Union nor the European Research Executive Agency can be held responsible for them.\footnotemark[3]
\footnotetext[3]{\href{https://cordis.europa.eu/project/id/101188037}{https://cordis.europa.eu/project/id/101188037}}

The authors acknowledge the use of Large Language Models for grammar checking, sentence rephrasing, and improving the readability of selected parts of the manuscript. All technical content, data analysis, interpretation of results, figures, and conclusions were developed, reviewed, and approved by the authors.
\bibliography{report} 
\bibliographystyle{spiebib} 

\end{document}